\documentclass[12pt,epsf]{article}

\usepackage{color}
\usepackage[dvips]{graphicx}
\usepackage{graphicx}
\usepackage{amsmath,amssymb}
\usepackage{epsfig}

\newcommand{\sfrac}[2]{\left(\frac{#1}{#2}\right)}

\begin{document}


\baselineskip 0.7cm

\begin{titlepage}

\begin{flushright}
UT-12-10\\
IPMU-12-0092\\
YITP-12-43
\end{flushright}

\vskip 1.35cm
\begin{center}
{\large \bf
Naive Dimensional Analysis in Holography}
\vskip 1.2cm
Ryoichi Nishio$^{1,2}$, Taizan Watari$^{1}$, Tsutomu T. Yanagida$^{1}$ and Kazuya Yonekura$^{3}$
\vskip 0.4cm

{\it $^1$ Kavli Institute for the Physics and Mathematics of 
the Universe (IPMU),\\ 
University of Tokyo, Chiba 277-8583, Japan\\
$^2$  Department of Physics, University of Tokyo,\\
    Tokyo 113-0033, Japan\\
$^3$ Yukawa Institute for Theoretical Physics, Kyoto University,\\ Kyoto 606-8502, Japan
}

\vskip 1.5cm

\abstract{ 
Naive dimensional analysis (NDA) is a widely used ansatz to estimate coupling constants among composite states emerging from dynamics of a strongly coupled gauge theory. However, the validity of NDA is still unclear because of the difficulty in calculating these quantities in strongly coupled theories.
We examine the NDA ansatz using gauge/string duality, by estimating glueball coupling constants from gravitational description.
The NDA scaling rule for coupling constants of some types of glueballs is 
verified and extended by both generic estimation and numerical
 calculations.
The scaling rule verified in this article can be applied to some class
 of quiver gauge theories as well, not just to gauge theories with a single gauge group $SU(N_c)$.
}
\end{center}
\end{titlepage}

\setcounter{page}{2}


\section{Introduction}
It is beneficial to theoretically understand parameters in the
low-energy effective theory of hadrons which emerges from a strong 
dynamics of a gauge theory.
Models beyond the standard model sometimes contain a strongly coupled 
sector (e.g. technicolor models, dynamical supersymmetry breaking
models, etc.), and the parameters of the low-energy effective theory of
hadrons can be observable or at least relevant to phenomenology.
Even within the standard model, the chiral Lagrangian is an effective 
theory of QCD.
Although the parameters of the effective theories such as masses and coupling constants should be determined in terms of parameters in the theories at short distance, it is often difficult to calculate them due to strong coupling.

However, there is an ansatz that is known to be reasonable to some
extent; it is called naive dimensional analysis (NDA)
\cite{Manohar:1983,Georgi:1986,Luty:1997fk,Cohen:1997rt},
which roughly guesses magnitude of coupling constants among hadrons.
In  the NDA ansatz, the effective action of glueballs is given by
\begin{align}
 S = \int d^4x \; \frac{N_c^2}{(4\pi\beta)^2} \; 
   {\cal L}(\phi(x), \partial_\mu, \Lambda_\text{NDA}),
\label{NDA_ansatz}
\end{align}
where $\Lambda_\text{NDA}$ is a parameter with mass dimension one and
fields collectively denoted by $\phi(x)$ represent glueballs.
All terms in ${\cal L}(\phi(x), \partial_\mu, \Lambda_\text{NDA})$ are assumed 
to have dimensionless coefficients of order unity (apart from an
appropriate power of $\Lambda_{\text{NDA}}$), and consequently, it follows that
 $\Lambda_\text{NDA}$ is the mass scale of hadrons (except for
 Nambu-Goldstone bosons).
The overall factor $N_c^2$ should be replaced with $N_c$ in the case of effective action of mesons \cite{Georgi:1992dw,Witten:1980sp}, so that the $N_c$ scaling rule in a large $N_c$ gauge theory is satisfied.
The essential point of the NDA ansatz is that the overall factor
contains $4\pi$ $(\simeq 13)$, which is sizable compared with unity;
this overall prefactor may not precisely be $N_c^2/(4\pi)^2$, but the
NDA ansatz assumes that the deviation---represented by a yet
undertermined factor $1/\beta^2$---is of order unity.  
After rescaling of $\phi(x)$ so that the $\phi(x)$ fields have canonically normalized kinetic terms, the coefficients of the interaction terms become of the forms, for example,
\begin{align}
 (\partial_\mu  \phi  \partial^\mu \phi) \left[\frac{\phi}{\Lambda_\text{NDA}}\sfrac{4\pi\beta}{N_c}\right]^I\times {\cal O}(1),
\quad
 \Lambda_\text{NDA}^2 \phi^2 \left[\frac{\phi}{\Lambda_\text{NDA}}\sfrac{4\pi\beta}{N_c}\right]^I\times {\cal O}(1)
\label{eq:example of coupling}
\end{align}
for arbitrary integers $I$ ($\ge 1$); any $(I+2)$-point coupling
constants will scale in $I$ as $(4\pi\beta/N_c)^{I}\times {\cal O}(1)$
in unit of $\Lambda_\text{NDA}$.  
There is an experimental support for the NDA ansatz (\ref{NDA_ansatz}, \ref{eq:example of coupling});
the coupling constant of three-pion-interaction $(\partial \pi)^2 \pi$ is $f_\pi^{-1}$ in the chiral Lagrangian, and it is reasonably close to the prediction of the NDA ansatz, $\frac{1}{\Lambda_\text{NDA}}\sfrac{4\pi}{N_c^{1/2}}$, when experimental values, $f_\pi\sim 100 $ MeV , $\Lambda_\text{NDA}\sim m_\rho \sim 1$ GeV and $N_c=3$ are substituted.

There are a couple of different ways to argue in favor of the NDA ansatz
(\ref{NDA_ansatz}) also from theoretical perspectives.
One of them is to require an ansatz of ``loop saturation'' in loop
calculation in the effective theory. 
The loop saturation ansatz requires that each loop contribution to an amplitude is
comparable to the tree level one, 
if the external momenta are taken to be of order $\Lambda_\text{NDA}$
and the UV-cutoff of loop momenta is also set to be $\Lambda_\text{NDA}$.
It follows from this ansatz that $(I+2)$-point coupling constants among arbitrary hadrons are ${\cal O}((4\pi)^{I})$.
However, the result seems to be incompatible with the $N_c$ scaling law, because the $(I+2)$-point coupling constants of glueballs 
should be proportional to $N_c^{-I}$ in a large $N_c$ gauge theory.
If one multiplies $(4\pi)^I$ by adequate powers of $N_c$ in order to satisfy the correct $N_c$ scaling law,
one obtains the NDA ansatz (\ref{NDA_ansatz}).

Another argument for the ansatz (\ref{NDA_ansatz}, \ref{eq:example of coupling}) is based on the large $N_c$ expansion.
Suppose that an operator ${\cal O}$ is a single trace operator such as $\text{Tr}(FF)$ which can create glueball states $|h_n\rangle$, $\langle \text{vac}| {\cal}O(x)| h_n ({\mathbf p})\rangle = f_n e^{ipx}$.
The $(I+2)$-point correlation function is given in planar limit by
\begin{align}
 \langle {\cal O}(p_1) {\cal O}(p_2) \dots {\cal O}(p_{I+2}) \rangle
=(2\pi)^4 \delta^4(\sum_i p_i) J_{I+2}(\lambda; {p_i}) \frac{N_c^2}{16\pi^2}.
\end{align}
If the 't Hooft coupling $\lambda= g_\text{YM}^2 N_c$ is small,
$J_{I+2}(\lambda; p_i)$ can be calculated perturbatively in the gauge theory and it is given schematically by
\begin{align}
 J_{I+2}(\lambda;p_i) \sim p^{d(I+2)+4} \times \left[
 {\cal O}(1)+ {\cal O}\sfrac{\lambda}{16\pi^2}+{\cal O}\sfrac{\lambda}{16\pi^2}^2+\dots
\right],
\end{align}
where $d$ is the dimension of the operator ${\cal O}(p)$.
The last factor $N_c^2/(16\pi)^2$ comes from two color loops of gluons and a loop factor.
For large $\lambda$, perturbation breaks down and we do not know the form of $J_{I+2}(\lambda; p_i)$.
If we knew $J_{I+2}(\lambda;p_i)$ for large $\lambda$,
we could read mass spectra $m_n$, decay constants $f_n$, and coupling constants $g_{n_1 \dots n_{I+2}}$ of glueballs
from the behavior of $J_{I+2}(\lambda;p_i)$ around mass poles $p_i^2 \rightarrow -m_{n_i}^2$ of each external momenta;
\begin{align}
J_2(\lambda; p_i) \frac{N_c^2}{16\pi^2}
 & \sim
 \frac{f_n^2}{p_1^2+m_n^2 }, &
 J_{I+2}(\lambda;p_i) \frac{N_c^2}{16\pi^2}
&\sim
\prod_{i=1}^{I+2} \sfrac{f_{n_{i}}}{p_{i}^2+ m^2_{n_{i}}} \times
g_{n_1 \dots n_{I+2}}.
\end{align}
If we assume that the unknown functions $J_{I+2}(\lambda;p_i)$ are of the form $p^{d(I+2)+4} \times {\cal O}(1)$ 
as in the case of small 't Hooft coupling,
 we may obtain
\begin{align}
 f_n &\sim m^{d+3} \sfrac{N_c}{4\pi} & g_{n_1 \dots n_{I+2}} \sim m^{-I+2} \times \sfrac{4\pi}{N_c}^{I},
\end{align}
where $m$ is the typical mass scale of glueballs.
The second expression is in accordance with the NDA ansatz (\ref{NDA_ansatz}).

Both of the above two arguments, however, are far from being a clear justification for the NDA ansatz.
The loop saturation in the first argument is an ansatz in itself.
Moreover, the apparent contradiction between the loop saturation and the $N_c$ scaling may cast doubt on the reasoning of loop saturation.
One may not be satisfied with the second argument either, because
the estimation for the Green functions $J_{I+2}(\lambda;p_i)$ from perturbative calculation has no justification for a large 't~Hooft coupling.

In this article, we study the NDA ansatz (\ref{NDA_ansatz}) by means of
gauge/string duality~\cite{Maldacena:1997re,Gubser:1998bc,Witten:1998qj}
for some types of scalar glueballs. 
Although the NDA has been checked only by experimental values of $f_\pi$
etc. in QCD so far, the gauge/string duality provides us with plenty of strongly coupled gauge theories
which are calculable in the gravity duals.
We leave a study of the NDA for other types of hadrons (such as glueballs with higher spin and flavor non-singlet mesons) for future work.
The rest of the paper is organized as follows.
In section 2, the glueball coupling constants are given in terms of 
the language of a string theory. 
They include overlap integrals which we cannot calculate analytically.
We will verify the NDA by giving a generic estimation for the overlap integrals.
In section 3, we confirm the estimation and the NDA scaling rule by numerical calculation in two specific gravity backgrounds.
In section 4, we briefly summarize our study.

Mass spectra and several three or four point couplings among composite states have already been studied by means of gauge/string duality in \cite{Sakai:2004cn, Sakai:2005yt, Mueck:2004qg, Hong:2004sa,Hong:2005np,Erlich:2005qh, Evans:2004ia} in several gravitational models.
Our aim in this article is to 
find a rule governing the behavior of large numbers of coupling constants,
not only three or four point couplings but generally $(I+2)$-point couplings.

Note also that we will not only verify the NDA ansatz (\ref{NDA_ansatz},
\ref{eq:example of coupling}), which is for $SU(N_c)$ gauge theories. 
In this paper, we derive an extended version of NDA scaling rule, which cover 
more general gauge theories (such as a class of quiver gauge theories
where the gauge group contains $p$ copies of $SU(N_c)$).
In the extended version of the NDA scaling rule, $N_c$ in the overall factor in (\ref{NDA_ansatz}) will be replaced with $\sqrt{a}$ multiplied by an ${\cal O}(1)$ factor, where $a$ is something like a ``degree of freedom'' of a gauge theory.

\section{Glueball Coupling Constants in Gauge/String Duality}
\label{sec:derivation}

In the gauge/string duality \cite{Maldacena:1997re,Gubser:1998bc,Witten:1998qj},
some gauge theories with a large number of color
 $N_c$ and a large 't Hooft coupling $\lambda=g_{YM}^2 N_c$ are dual to string theories on warped spaces.
A gravity dual of a four-dimensional {\it conformal} field theory is the type-IIB string theory on a spacetime $AdS_5 \times W$ with a metric
\begin{align}
 ds^2 = R^2/z^2 (\eta_{\mu\nu}dx^\mu dx^\nu + dz^2) + R^2ds_W^2,\label{181343_4Mar12}\quad 0 < z < \infty,
\end{align}
where $W$ is a five-dimensional Einstein manifold. 
An infinite number of examples of $W$ are known, with each manifold $W$
corresponding to a conformal field theory.
For example, the Type IIB string with $W = S^5$ is dual to ${\cal N}=4$
$SU(N_c)$ super Yang-Mills theory \cite{Maldacena:1997re}, and a choice 
$W = T^{1,1}$ corresponds to an ${\cal N}=1$ $SU(N_c)\times SU(N_c)$ gauge theory with four chiral multiplets in the bifundamental representation \cite{Klebanov:1998hh}.
In references \cite{Gauntlett:2004yd, Benvenuti:2004dy},
one can find an infinite number of examples of gravity background
geometries with $W=Y^{p,q}$, which correspond to 
$(SU(N_c))^p = SU(N_c) \times SU(N_c) \times \cdots \times SU(N_c)$ ($p$
factors of $SU(N_c)$) quiver gauge theories with several chiral multiplets in the bifundamental representations between two of the $p$ $SU(N_c)$ groups.

Gravity duals of {\it confining} gauge theories have also been constructed \cite{Witten:1998zw,Polchinski:2000uf,Klebanov:2000hb, Maldacena:2000yy, Butti:2004pk}.
In most of them, the metric background in the string frame may be written without loss of generality as,
\begin{align}
 ds^2 = (f(z))^{-2}(\eta_{\mu\nu}dx^\mu dx^\nu+ dz^2)+R^2 ds_{W_z}^2,
\label{eq:general geometry}
\end{align}
with an appropriate definition of the $z$-coordinate.
The warp factor $R^2/z^2$ in (\ref{181343_4Mar12}) for the case of conformal theories  
is replaced
by a more general form $(f(z))^{-2}$.
The internal manifold $W_z$ can also be dependent on $z$,
and the other supergravity fields generally have nontrivial background.
Although details of the background are model dependent,
most of the gravitational models which are dual to confining gauge theories have
at least one thing in common;
the radial coordinate terminates at finite value\footnote{The necessary and sufficient condition for confinement of {\it heavy quarks}
are that $(f(z))^{-2}$ has a nonzero minimal value \cite{Kinar:1998vq},
but this condition necessarily provides neither a finite $z_\text{max}$ nor discrete spectrum of glueballs.
However, in most of gravitational models with confinement of heavy quarks, 
$z_\text{max}$ is also finite and $(f(z))^{-2}$ has its finite minimal value at $z=z_\text{max}$. } 
$z=z_\text{max}$.
Normalizable wavefunctions of the fluctuations of supergravity fields
around their backgrounds become discrete like cavity modes when the
range of the radial coordinate ($0 < z < z_\text{max}$) is finite.
Each composite state, i.e. glueball, in a gauge theory is described by each normalizable wavefunction in the dual gravity theory. 
The quantity $z_\text{max}$ corresponds to the mass scale of glueballs in the gauge theory.

The four dimensional effective action of glueballs is obtained by dimensional reduction:
expanding the fields in the ten dimensional type-IIB supergravity action 
in terms of the complete set of wavefunctions on the radial and compact directions
and integrating over those directions.
We can carry out this procedure by two steps of dimensional reduction;
the first step is to construct a five dimensional supergravity action by integrating the complete set on the five dimensional compact manifold $W_z$, and the second step is to obtain the four dimensional effective action of glueballs by integrating the complete set on the radial $z$-coordinate.

\subsection{An Intermediate Description in Five Dimensions}
In order to examine how glueball couplings are controlled in gauge/string duality, we start from dimensional reduction on five dimensional internal manifold in the case of the conformal background (\ref{181343_4Mar12}). 
It is useful to study the conformal case first, before going to the confining case (\ref{eq:general geometry}), because
the conformal background (\ref{181343_4Mar12}) approximates the confining background (\ref{eq:general geometry}) to some extent, except for deep IR-region $z\sim z_\text{max}$.

In the case of the conformal background $AdS_5\times W$ (\ref{181343_4Mar12}), dilaton has constant background, $e^{\langle \Phi\rangle}=g_s$, and the self dual five form $F_5 = (1+*){\cal F}_5$ has background
 ${\cal F}_5= \frac{(2\pi \sqrt{\alpha'})^4}{\text{Vol}(W)} N_c \; \text{vol}(W)$,
where $\text{vol}(W)$ is the volume form of $W$ and $\text{Vol}(W)$ is its volume; $\text{Vol}(W)= \int_W \text{vol}(W)$. 
The backgrounds of the other fields, $C_0$, $B_2$ and $C_2$, are constant, which we assume to be zero for simplicity.
After dimensional reduction on the five dimensional compact manifold $W$,
the supergravity action using the Einstein frame metric becomes
\begin{align}
S=\frac{(R^5 \text{Vol}(W)) \times R^3}{2\kappa_{10}^2 g_s^2}  \int d^4x \int^\infty_0 \frac{dz}{z^3}
 \left( -\frac{1}{2}((\partial \phi)^2+(\partial_z \phi)^2)-\frac{1}{2} e^{2\phi}((\partial c)^2+(\partial_z c)^2) \right)+\dots,\label{eq:cftaction}
\end{align}
where ${2\kappa_{10}^2}=(2\pi)^7 \alpha'^4$ is the gravitational coupling constant of ten dimensional type IIB supergravity.
Here we only keep track of $\phi(x,z)$ and $c(x,z)$ fields in five dimensions\footnote{
The terms $+\cdots$ in (\ref{eq:cftaction}) represent terms including the fluctuations other than $\phi$ and $c$. 
However, it includes no
mixing in the bilinear terms between $\phi$ or $c$ and the other fluctuations in the case of the $AdS$ background considered here.
On the other hand, the terms $+\cdots$ also includes interaction terms
among $\phi$ or $c$ and the other fluctuations such as $(\partial_m \phi)(\partial_n \phi) h^{mn}$, where $h_{mn}$ is a fluctuation of the metric.
We are keeping only the interaction terms among $\phi$ and $c$. 
},
which correspond to fluctuations of $\Phi$ (dilaton) and $C_0$ (RR scalar) with a constant profile on $W$, respectively. 
The AdS radius $R$ is given by
\begin{align}
 R^4 = 4\pi^4 g_s \alpha'^2 \frac{1}{\text{Vol}(W)}= 4\pi g_s \alpha'^2 \frac{\text{Vol}(S^5)}{\text{Vol}(W)}.
\label{eq:AdSradius}
\end{align}
Then the overall factor is rewritten as 
\begin{align}
\frac{1}{2\kappa_{10}^2 g_s^2} \times (R^5 \text{Vol}(W)) \times R^3 =
 \frac{4a}{8\pi^2},
\end{align}
with the definitions  
\begin{align}
 4a \equiv p' N_c^2, \quad p' \equiv \frac{\text{Vol}(S^5)}{\text{Vol}(W)}.\label{eq:def of a}
\end{align}

The supergravity action (\ref{eq:cftaction})
becomes an expression that is interesting in the context of the NDA, 
by canonically rescaling the fields,
\begin{align}
\notag
 S=
\int d^4x \int^\infty_0 \frac{dz}{z^3} 
  \left[ -\frac{1}{2}((\partial \phi')^2+(\partial_z \phi')^2)-\frac{1}{2}((\partial c')^2+(\partial_z c')^2) \right.\\
\left. - \sum_{I=1}^\infty\sfrac{4\pi}{\sqrt{2a}}^I \frac{1}{ I! 2!} (\phi')^I ((\partial c')^2+(\partial_z c')^2)\right].\label{CFT5dimaction}
\end{align}
Here, $\phi'=\phi - \langle \Phi\rangle$ and $c'=g_s c$.
The second line plays a role of interaction terms, which comes from the exponential dilaton factor $e^{2\phi}$.
Besides combinatoric factors, the $(I+2)$-point interaction terms of the five dimensional fields $\phi'$ and $c'$ have coefficients $((4\pi)/\sqrt{2a})^{I}$. Although these coefficients are not glueballs coupling constant themselves, 
this expression (\ref{CFT5dimaction}) is already quite indicative that 
the coupling constants of glueballs may show the behavior predicted by NDA (\ref{NDA_ansatz}).

We saw that the parameter $a$ in (\ref{eq:def of a}) controls the interactions of the closed string fields.
The parameter $a$ corresponds to the central charge $a$ of the conformal field theory\footnote{For a review, see e.g. \cite{Duff:1993wm}.
In conformal field theory, another central charge $c$ is also defined. Although $a$ and $c$ are different in generic conformal field theories,
one has the relation $a=c$ when the theory has a weakly coupled gravity description as in the case we are studying in this paper.
We just use the symbol $a$ throughout this paper.
}
which is dual to  $AdS_5\times W$ \cite{Gubser:1998vd,Henningson:1998gx}.
The central charge $a$ represents something like degrees of freedom of the conformal field theory.\footnote{For example,
in a free massless field theory with $n_s$ scalars, $n_f$ Weyl fermions and $n_v$ vector bosons, one has~\cite{Duff:1993wm}
$a=\frac{1}{360}\left(n_s+\frac{11}{2}n_f+62n_v \right)$ and
$c= \frac{1}{120} \left(n_s+3n_f+12n_v \right)$.
}
For example, the property $a \propto N_c^2$ can be understood because degrees of freedom of gluons and other fields in the adjoint representation of $SU(N_c)$ gauge group are ${\cal O}(N_c^2)$.
Moreover, in the gravity dual of $(SU(N_c))^p$ quiver gauge theory, $p'$ defined in (\ref{eq:def of a}) is ${\cal O}(p)$ \cite{Gauntlett:2004yd,Benvenuti:2004dy}, 
so we have $a \sim  pN_c^2$. This is also reasonable for the interpretation that $a$ is roughly degrees of freedom of the conformal field theory because degrees of freedom of gauge theory with the gauge group $SU(N)^p$ should be about $pN_c^2$.
We also find that $a$ has a lower bound $4a \ge N_c^2$~\cite{Gauntlett:2006vf}
because it is proven that $p'$ has a lower bound $p'\ge 1$ in general Einstein manifold.
Therefore, the parameter $a$ makes the couplings in (\ref{CFT5dimaction}) always smaller than $(4\pi)^I$.

\subsection{Effective Action in Four Dimensions}\label{sec:effectiveaction}
In order to obtain the effective action of glueballs on the four dimensional spacetime,
we also have to integrate the fifth coordinate $z$.
The supergravity action on $AdS_5$ (\ref{eq:cftaction},
\ref{CFT5dimaction}) is no longer suitable for this purpose, because
conformal theories do not give rise to a discrete hadron spectrum.
We start to consider supergravity on confining geometries
(\ref{eq:general geometry}) instead, and describe coupling constants of glueballs.

Let us first consider a theory which is asymptotically conformal in the UV region.
In this case, only minor modification to the five-dimensional description (\ref{eq:cftaction}, \ref{CFT5dimaction}) is necessary. 
In the gravity dual language, $(f(z))^{-2} \simeq (R/z)^2$ in the $z\rightarrow  0$ limit, and the central charge $a$ in the UV limit is defined by (\ref{eq:def of a}) by using $\text{Vol}(W_z)$ at $z=0$.
On such a geometry, with the same rescaling of the fields as in (\ref{CFT5dimaction}),
the five dimensional supergravity action is written as
\begin{align}
 S=
\int d^4x \int^{z_\text{max}}_0 \frac{dz}{z^3} Y(z)
  \left[ -\frac{1}{2}((\partial \phi')^2+(\partial_z \phi')^2)-\frac{1}{2}((\partial c')^2+(\partial_z c')^2) \right.\notag \\
\left. -\frac{1}{2! I!} \sum_{I=1}^\infty \sfrac{4\pi}{\sqrt{2a}}^I (\phi')^I ((\partial c')^2+(\partial_z c')^2)\right]+\dots,\label{5dimactionconfining}
\end{align}
where $Y(z)$ is a dimensionless function defined as
\begin{align}
 Y(z)= \frac{(f(z))^{-3}}{(R/z)^3}\frac{\text{Vol}(W_z)}{\text{Vol}{(W_{z=0})}},
\end{align}
which is unity in the case of conformal geometries.
In the $z\rightarrow 0$ (UV) limit, $Y(z)\rightarrow 1$, and the 
integrand of (\ref{5dimactionconfining}) becomes identical to 
one of (\ref{CFT5dimaction}).

In the rest of this article, we omit the terms in $+\cdots$ in (\ref{5dimactionconfining}).
In general confining geometry, dilaton, RR scalar and the 3-form flux $H_3$ and $F_3$ would also have nontrivial background. 
In this case $\phi'(x,z)$ and $c'(x,z)$
would have mixings with other string fields such as the metric. 
We neglect this technical complexity in this article.
We expect that such details of the IR background will affect coupling
constants of glueballs at most by ${\cal O}(1)$ factors, and hence they are
unessential in trying to verify the NDA ansatz, whose predictions 
always come with uncertainty of order unity.

We denote four dimensional glueball fields of the $n$-th excited modes by $\tilde \phi_n(x)$ and $\tilde c_n(x)$, which are created by operators dual to five-dimensional supergravity fields $\phi'(x,z)$ and $c'(x,z)$, respectively. We assign mass dimension one for glueball fields $\tilde \phi_n(x)$ and $\tilde c_n(x)$
just as usual for canonically normalized scalar fields in four dimensional field theory.
The five dimensional fields $\phi'(x,z)$ and $c'(x,z)$ are decomposed
into independent modes in four-dimensions, each one of which corresponds to
a normalizable wavefunction $\psi_n(z)$;
\begin{align}
 \phi'(x,z) = \sum_{m=1}^\infty \psi_m(z) \tilde \phi_m(x),
\quad
 c'(x,z) = \sum_{n=1}^\infty \psi_n(z) \tilde c_n(x).\label{224151_16Mar12}
\end{align}
The normalizable modes $\psi_n(z)$ are defined as the solutions of the eigen-equation given by
\begin{align}
z^3Y^{-1}(z) \partial_z \left(z^{-3}Y(z) \partial_z \psi_n(z) \right) =  m_n^2 \psi_n(z),\label{174636_19Mar12}
\end{align}
with the normalization condition
\begin{align}
 \int_0^{z_\text{max}} \frac{dz}{z^3}Y(z) \psi_n(z) \psi_m(z)= \delta_{nm},\label{224759_16Mar12}
\end{align}
which makes all the fields $\tilde \phi_n(x)$ and $\tilde c_n(x)$
canonically normalized in the 4D effective action.
The modes $\psi_n(z)$ satisfies an appropriate IR-boundary condition which
is imposed so that the field configuration in ten dimensional spacetime should be smooth
at $z=z_\text{max}$. 
The eigenvalue $m_n$ is the mass of glueballs\footnote{
The normalizable wavefunctions $\psi_n(z)$ and mass spectra $m_n$ are common in $\phi'$ and $c'$ in our study.
This is because we are ignoring any kind of nontrivial vacuum expectation values and 
effects of mixing for the fields of supergravity, 
and therefore equations of motion for $\phi'$ and $c'$ become the same.
} $\tilde \phi_n$, $\tilde c_n$.
The effective action of glueballs is obtained by substituting (\ref{224151_16Mar12}) to (\ref{5dimactionconfining}).

The interaction part of the effective Lagrangian  ${\cal L}_\text{int}$ 
includes $(I+2)$-point interaction terms 
which consist of two types of couplings; 
one is of the form $\tilde \phi_m^I(\partial \tilde c_n)^2$,
 and the other $\tilde \phi_m^I \tilde c_n^2$ without derivatives:
\begin{align}
 {\cal L}_\text{int}=- \sum_{I=1}^\infty
 \sum_{n_1 n_2 m_1 \dots m_I}&\left[
 a^{(I)}_{n_1 n_2, m_1\dots m_I} 
\frac{\Lambda^{-I}_\text{NDA}}{2!I!} \tilde \phi_{m_1} \dots\tilde  \phi_{m_I} (\partial_\mu \tilde c_{n_1}) (\partial^{\mu} \tilde c_{n_2})
\right.\notag \\
& \left.+
b^{(I)}_{n_1 n_2, m_1\dots m_I} \frac{\Lambda^{2-I}_\text{NDA}}{2!I!} \tilde  \phi_{m_1} \dots\tilde  \phi_{m_I}  \tilde c_{n_1}\tilde  c_{n_2}
\right].\label{183939_19Mar12}
\end{align}
The coupling constants $a^{(I)}_{n_1n_2m_1\dots m_I}$ and $b^{(I)}_{n_1 n_2 \dots m_1 \dots m_I}$ are given by the following overlap integral of normalizable wavefunctions;
\begin{align}
 a^{(I)}_{n_1n_2 m_1\dots m_I} &= 
 \sfrac{4\pi}{\sqrt{2a}}^I \times
    \Lambda^I_\text{NDA}\int_0^{z_\text{max}}\frac{dz}{z^3} Y(z) \psi_{n_1} \psi_{n_2} \psi_{m_1} \dots \psi_{m_I},\label{a}\\
 b^{(I)}_{n_1n_2 m_1\dots m_I} &= 
 \sfrac{4\pi}{\sqrt{2a}}^I \times
 \Lambda^{I-2}_\text{NDA}
\int_0^{z_\text{max}}\frac{dz}{z^3} Y(z)(\partial_z\psi_{n_1}) (\partial_z \psi_{n_2}) \psi_{m_1} \dots \psi_{m_I}.\label{b}
\end{align}
These coupling constants  have been made dimensionless
by multiplying appropriate powers of a parameter $\Lambda_\text{NDA}$ with mass dimension one.
When we choose the mass scale $\Lambda_\text{NDA}$ as the mass of the lightest glueball mass, these coupling constants are expected to be ${\cal O}(4\pi\beta/N_c)$ by the NDA ansatz (\ref{eq:example of coupling}).

We need to estimate the overlap integrals of (\ref{a}, \ref{b}) in order to
examine whether there exists a rule governing the coupling constants of
four dimensional effective theory like the NDA ansatz or not.
The prefactor $(4\pi/\sqrt{2a})^{I}$ is identical to the coefficient of 
$(I+2)$-point interaction terms of (\ref{CFT5dimaction}), and
is just determined only from conformal region in UV,
whereas the remaining overlap integrals are dependent on the detail of the IR geometry.
The estimation of the overlap integrals unavoidably requires numerical calculations in each geometry and they will be the subject of section 3.
Before numerical calculation, however, we will give a crude estimation
independently of the detail of geometries.

We may roughly estimate the overlap integrals  for low excited modes and not too large $I$ by approximating the integrand as a constant value in the IR region.
 On general UV-conformal geometry,
 normalizable wavefunctions behaves as $z^\Delta$ ($\Delta$ is the conformal dimension and $\Delta=4$ for $\phi(x,z)$ and $c(x,z)$) in the small $z$ region, so the small $z$ region have only small contribution to the overlap integrals.
They oscillate with relatively large amplitudes in the IR region $z\sim {\cal O}(z_\text{max})$, say, $z \gtrsim
 z_\text{max}/2$.
The normalizable wavefunctions 
of the first few excited modes have only a small number of nodes.
Then it might be justified to approximate the integrand by a typical
constant value in the IR region. 
Using the value of the integrand at around 
$z \sim \gamma z_\text{max}~(\gamma \sim {\cal O}(1))$ as the typical value, the overlap integrals are estimated as
\begin{align}
  |a^{(I)}_{n_1n_2 m_1\dots m_I}| &\sim
\sfrac{4\pi}{\sqrt{2a}}^{I}
 \Lambda^I_\text{NDA} z_\text{max}^{-2} \left[Y(z) \psi_{n_1} \psi_{n_2}
 \psi_{m_1} \dots \psi_{m_I}\right]_{z\sim \gamma z_\text{max}}, \label{195513_19Mar12}
 \\
  |b^{(I)}_{n_1n_2 m_1\dots m_I}| &\sim
 \sfrac{4\pi}{\sqrt{2a}}^I \Lambda^{I-2}_\text{NDA}z_\text{max}^{-2} \left[Y(z) (\partial_z\psi_{n_1}) (\partial_z\psi_{n_2}) \psi_{m_1} \dots \psi_{m_I}\right]_{z\sim \gamma z_\text{max}}.      \label{195521_19Mar12}
 \end{align}
Applying the same approximation for bilinear terms, we also obtain
  \begin{align}
  \psi_n(z\sim \gamma z_\text{max})\sim
 \left[ z_\text{max}^{-2} Y(z\sim \gamma z_\text{max})\right]^{-1/2},\quad   \partial_z \psi_n(z\sim \gamma z_\text{max})\sim m \psi_n(z\sim \gamma z_\text{max}).\label{195506_19Mar12}
 \end{align}
We are assuming that $n_i, m_j$ and $I$ are not large, because the integrand would oscillate rapidly if they were large and the above estimation would break down.
Here, $\gamma z_\text{max}$ is a typical value of $z$ around which the integrand is peaked, and should be near the IR boundary, say, $\gamma \sim 1/2$, and
$m$ is a typical mass of the low excited modes (say, $m=m_1$).
 Substituting (\ref{195506_19Mar12}) to (\ref{195513_19Mar12}, \ref{195521_19Mar12}),  we obtain 
 \begin{align}
   |a^{(I)}_{n_1n_2 m_1\dots m_I}|&\sim 
 \left[
 \sfrac{4\pi}{\sqrt{2a}}
  \Lambda_\text{NDA}z_\text{max} \left( Y(z\sim \gamma z_\text{max})\right)^{-1/2}  \right]^{I},\label{200404_19Mar12}
 \\
   |b^{(I)}_{n_1n_2 m_1\dots m_I}|&\sim \sfrac{m}{\Lambda_\text{NDA}}^2
 \left[
 \sfrac{4\pi}{\sqrt{2a}}
  \Lambda_\text{NDA} z_\text{max} \left( Y(z\sim \gamma z_\text{max})\right)^{-1/2}
 \right]^{I}.\label{200411_19Mar12}
 \end{align}
 The reason $b^{(I)}$ has the factor of $(m/\Lambda_\text{NDA})^2$ is that (\ref{195521_19Mar12}) contains two $(\partial_z \psi_n)$'s.
 
With these crude approximations,
 one  can find that there is a scaling rule for coupling constants like  the NDA ansatz (\ref{NDA_ansatz}).
When we choose $\Lambda_\text{NDA}$ as the typical mass of glueballs, 
\begin{align}
 \Lambda_\text{NDA}&= m, \label{eq:choice of LambdaNDA}
\end{align}
both of the $(I+2)$-point coupling constants 
are estimated as $I$-th powers of the same factor,
\begin{align}
   |a^{(I)}_{n_1n_2 m_1\dots m_I}|,  |b^{(I)}_{n_1n_2 m_1\dots m_I}|
&\sim 
\left[
\sfrac{4\pi}{\sqrt{2a}} m z_\text{max}(Y(z\sim \gamma z_\text{max}))^{-1/2}
\right]^I
\notag 
\\
&\sim
 \left[
 \sfrac{4\pi}{\sqrt{2a}} (m z_\text{max})
     \right]^{I},\label{005428_23Apr12}
\end{align}
where we have also used an approximation $Y(z\sim \gamma z_\text{max})\sim 1$ in the second line.
The factor $(m z_\text{max})$
is expected to be ${\cal O}(1)$ because $1/z_\text{max}$ corresponds to confinement scale 
of dual gauge theory, but this factor turns out to be slightly large in numerical calculations as we will see in the next section.
Eq. (\ref{005428_23Apr12}) with the choice of (\ref{eq:choice of LambdaNDA}) is just the same as what the NDA predicts (\ref{NDA_ansatz}), with the identification of the NDA scaling factor
\begin{align}
 \frac{4\pi\beta}{N_c}& \sim  \left[
 \sfrac{4\pi}{\sqrt{2a}} (m z_\text{max})
  \right]
 .\label{Eq:our statement}
\end{align}

The NDA scaling rule shown in (\ref{Eq:our statement}) has been generalized
from the original NDA (\ref{NDA_ansatz});
the factor $N_c/\beta$ in the original NDA rule (\ref{NDA_ansatz})
is generalized into $\sqrt{2a} /(m z_\text{max})$.
The identification is natural in an $SU(N_c)$ gauge theory because the factor $\sqrt{2a} /(m z_\text{max})$ is  actually
${\cal O}(N_c)$, which corresponds to the assumption of the original NDA that $\beta$ is ${\cal O}(1)$.
However, Eq.(\ref{Eq:our statement}) implies that $\beta$ 
can take an arbitrarily small value, because $4a/N_c^2=p'$ can be arbitrarily large, for example, in a quiver gauge theory $(SU(N_c))^p$
with arbitrarily large $p$. On the other hand, we also find that $\beta$ has an upper bound of ${\cal O}(1)$ value because $p'$ is bounded as $p' \ge 1$.

So far, we have only focused on UV-conformal theories, but 
it is possible to extend the derivation above in order to cover theories with weakly running couplings even in the UV-limit.
To do this, note that the $z$-dependent function $a(z)$ can be defined in gravity side even in non-conformal theories\footnote{
In a gauge theory which has both IR and UV fixed points, the central charge $a$ (which is equal to $c$ in supergravity approximation) has a relation $a_\text{UV}\ge a_\text{IR}$ (see \cite{Komargodski:2011vj} and references therein).
The $z$-dependent function $a(z)$ in (\ref{eq:a(z)}) is defined so that
it decreases monotonically as the holographic coordinate $z$ is increased~\cite{Freedman:1999gp}.
} \cite{Freedman:1999gp}:
\begin{align}
 a(z)=\frac{2\pi^2 R^5}{(2\pi)^7g_s^2\alpha'^4} \left(\frac{d}{dz}(f(z)(\text{Vol}(W_z))^{-1/3})\right)^{-3}.
\label{eq:a(z)}
\end{align}
One can see that $a(z)$ approaches the value $a$ defined in (\ref{eq:def of a}) when the geometry approaches $AdS_5\times W$.
In a theory which has a weakly running coupling in UV,
the function $a(z)$ varies only slowly in $z$, although it may vary
rapidly in the deep IR region $z\sim z_\text{max}$.
Then the above NDA scaling rule (\ref{Eq:our statement})
is expected to be valid also in such theories by replacing $a$ 
with $\bar{a}$, a typical value of $a(z)$;
\begin{align}
  \frac{4\pi\beta}{N_c}& \sim \left[
 \sfrac{4\pi}{\sqrt{2\bar{a}}} (m z_\text{max})
  \right].\label{Eq:our statement2}
\end{align}
Here, $\bar{a} = a(z\sim \gamma z_\text{max})$,
and $z\sim \gamma z_\text{max}$ is a typical value of $z$ around which
the overlap integral is dominated. The typical value $\bar{a}$ has an ambiguity which is at most ${\cal O}(1)$ factor depending on the choice of $\gamma$ because $a(z)$ is slowly varying, 
and such uncertainty is already expected in the estimation of this section.

\section{Numerical Check}

In the previous section, we have verified the NDA scaling rule by 
a crude estimation of the overlap integrals.
However, it is difficult to estimate the error within that crude estimation.
If the error in the estimation of (\ref{Eq:our statement}) were greater than the factor of order $4\pi$,
that error would be too large to claim that the NDA really holds. 
In this section we want to confirm the NDA scaling rule (\ref{Eq:our statement}, \ref{Eq:our statement2})
by calculating coupling constants numerically, and checking that they
are approximately given by the expected value (\ref{005428_23Apr12}) within the range of ${\cal O}(1)$ factors.
We will perform numerical calculations in two specific gravity models.
 Our results are in considerably good agreement with (\ref{Eq:our statement}, \ref{Eq:our statement2}). 

\subsection{Hard Wall Model}
\label{ssec:HW}

First, we estimate the glueball couplings in ``hard wall'' model \cite{Polchinski:2001tt}, which is regarded as the simplest toy model of IR-confining and UV-conformal gauge theories. 
The hard wall model simply introduces IR-boundary $z = z_\text{max}$ to the geometry (\ref{181343_4Mar12}) which is dual to a conformal field theory.
Thus, the 5D action of this model is given by (\ref{5dimactionconfining}) with $Y(z)=1.$
In this article, we choose Dirichlet boundary condition at the IR
boundary on the wavefunctions, i.e., $\psi_n(z_\text{max})=0$ (just for simplicity).

In the hard wall model, the mass spectra and normalizable wavefunctions
are well-known. The wavefunction $\psi_n(z)$ and mass eigenvalue $m_n$
of the $n$-th excited state are 
\begin{align}
 \psi_n(z) 
&=
  \frac{\sqrt{2}z^2}{z_\text{max}}
\frac{J_2(z m_n)} {J'_2(j_{2,n})}
,\quad m_n =j_{2,n}/z_\text{max},
\label{eq:wf in hw}
\end{align}
where $J_2(x)$ is the Bessel function of the second order, and 
$j_{2,n}$ the $n$-th zero of $J_2(x)$; 
$j_{2,1}\simeq 5.1$, $j_{2,2}\simeq 8.4$, $j_{2,3}\simeq 11$.  
The dimensionless coefficients of the $(I+2)$-point couplings, 
$a^{(I)}_{n_1 n_2 m_1 \dots m_I}$ and $b^{(I)}_{n_1 n_2 m_1 \dots m_I}$,  
can be calculated by using these wavefunctions in (\ref{a}, \ref{b}).

We calculated these coefficients numerically, and obtained geometric means 
$a^{(I)}_{\text {typ}}$ and $b^{(I)}_{\text{typ}}$ of the ensemble 
of the $(I+2)$-point coupling constants involving lower excited 
states.\footnote{For the numerical results of $a^{(I)}_{\text{typ}}$,
$\sigma^{(I)}_{\ln a}$, $b^{(I)}_{\text{typ}}$ and $\sigma^{(I)}_{\ln
b}$, we used an ensemble of $|a^{(I)}_{n_1 n_2 m_1 \dots m_I}|$ and
$|b^{(I)}_{n_1 n_2 m_1\dots m_I}|$ for a given value of $I$, 
with $1\le n_1\le n_2\le m_1 \le \dots \le m_I \le 3$ and 
$1\le n_1\le n_2\le 3,\; 1\le m_1 \le \dots \le m_I \le 3$.} 
The result shows that the one of the typical values of the $(I+2)$-point 
coupling constants, $a^{(I)}_{\text{typ}}$, remains almost an $I$-independent 
constant of order unity ($c \simeq 0.3\mbox{--}0.4$), after an
appropriate scaling behavior is factored out: 
\begin{equation}
 a^{(I)}_{\text{typ}} \simeq c \times 
   \left(\frac{4\pi}{\sqrt{2a}} (\Lambda_{\rm NDA} z_{\rm max}) \times
    0.8 \right)^I. 
\label{eq:HW-scaling-atyp}
\end{equation}
See Table \ref{HWresult}. A similar result was also obtained for 
the other typical value $b^{(I)}_{\text{typ}}$ of the $(I+2)$-point 
coupling constants when we take $\Lambda_\text{NDA}=m_1\simeq 5.1/z_\text{max}$.

This numerical result shows that the crude estimates in the previous
section are quite accurate on average, and there is no question about 
the scaling behavior of the $(I+2)$-point coupling constants now. 
It is worthwhile to note, further, that the scaling factor in
(\ref{Eq:our statement}) in the scaling rule we established is somewhat 
larger---by a factor of $m_1 z_{\max} \simeq 5.1$ or so in the hard wall
model---than what was expected in the naive dimensional 
analysis.\footnote{The difference between $(4\pi)$ and $(4\pi) \times 5.1$ may 
yield different phenomenology, when the scaling rule is
applied to physics beyond the Standard Model. }

Although we have seen so far that the geometric {\it means} of $(I+2)$-point
coupling constants of hadrons, $a^{(I)}_{\rm typ}$ and $b^{(I)}_{\rm typ}$, 
follow the scaling rule, such a rule will be of little value if the individual
coupling constants $a^{(I)}_{n_1 n_2 m_1 \dots m_I}$ and 
$b^{(I)}_{n_1 n_2 m_1 \dots m_I}$ take values wildly different from
their averages. 
Our numerical study shows that 
the typical range of $|a^{(I)}_{n_1 n_2 m_1 \dots m_I}|$ is from 
$a^{(I)}_{\rm typ} \times [e^{- \sigma^{(I)}_{\ln a}}]$ to 
$a^{(I)}_{\rm typ} \times [e^{+ \sigma^{(I)}_{\ln a}}]$, with 
$[e^{\sigma^{(I)}_{\ln a}}] \simeq 2.$ in the hard wall model. 
Similarly, it turns out that $[e^{\sigma^{(I)}_{\ln b}}] \simeq 3$.
Thus, the coupling constants are typically within the factor of 
$2 \sim 3$ from their geometric means.
With the scaling factor $(4\pi) \times (\Lambda_{\rm NDA} z_{\rm max})$
being much larger than the typical difference among the 
individual couplings $[e^{\sigma^{(I)}_{\ln}}]$, we see that 
the scaling rule of the averaged value (\ref{eq:HW-scaling-atyp}) 
contains valuable information (albeit statistical) on individual 
$(I+2)$-point coupling constants.

\begin{table}[tbp]
 \caption{ \label{HWresult} Geometric means
$a^{(I)}_\text{typ}$ and  $b^{(I)}_\text{typ}$ and 
the standard deviations $\sigma^{(I)}_{\ln a}$ and 
$\sigma^{(I)}_{\ln b}$ calculated numerically in the hard wall model. 
The standard deviation of $\ln | a^{(I)}_{n_1n_2 m_1 \dots m_I}|$'s, 
that is, $\sigma^{(I)}_{\ln a}$, is presented in the 3rd column in the form of 
$\ln [\;  \exp (\sigma^{(I)}_{\ln a}) \; ]$, so that  
the range of typical values of $|a^{(I)}_{n_1 n_2 m_1 \dots m_I}|$ 
can be easily read out.
}
 \begin{center}
 \begin{tabular}[c]{|c|c|c|c|c|}
 \hline 
 &
\hspace{-1.8mm}$\displaystyle{a^{(I)}_\text{typ}\left(\frac{4\pi}{\sqrt{2a}} \Lambda_\text{NDA}z_\text{max} \cdot 0.8 \right)^{-I}}$\hspace{-1mm}
 &
\hspace{-1mm} $\sigma^{(I)}_{\ln a}$ 
 &
 $\hspace{-1.8mm}\displaystyle{b^{(I)}_\text{typ}\left(\frac{4\pi}{\sqrt{2a}}
	      \Lambda_\text{NDA}z_\text{max} \cdot 0.8
	      \right)^{-I}}$\hspace{-1mm}& 
$\sigma^{(I)}_{\ln b}$ 
\\
\hline
$I=1$ &
0.3 &
$\ln [2.]$
&
$\sfrac{m_1}{\Lambda_\text{NDA}}^2 \times 0.4$
&
$\ln [3.]$
\\
$I=2$ &
0.3 &
$\ln [2.]$
&
$\sfrac{m_1}{\Lambda_\text{NDA}}^2 \times 0.3$
&
$\ln [3.]$
\\
$I=3$ &
0.4 &
$\ln [2.]$
&
$\sfrac{m_1}{\Lambda_\text{NDA}}^2 \times 0.2$
&
$\ln [3.]$
\\
$I=4$ &
0.4 &
$\ln [2.]$
&
$\sfrac{m_1}{\Lambda_\text{NDA}}^2 \times 0.4$
&
$\ln [4.]$
\\\hline
 \end{tabular}
 \end{center}
 \end{table}

\subsection{Klebanov-Strassler Metric}
\label{ssec:KS}

One of the flaws of the hard wall model is that the IR-region of the
geometry is very ad hoc; the IR-boundary $z_\text{max}$ is introduced by hand.
Such a crude treatment is meant only to be a simplest toy model
imaginable that implements the two essential ingredients of IR confining 
models: i) finite range of the holographic radius, 
${}^{\exists} z_{\rm max} \geq z \geq 0$, and ii) existence of the
minimal value of the warped factor $f^{-2}(z)$.
It is not meant at all to be a faithful (and hence stable) solution of the 
equation of motion of the Type IIB string theory. 
 
In a full solution of equations of motion of supergravity, however, 
the IR boundary $z = z_{\rm max}$ is not a singularity of the background 
geometry; the spacetime geometry is smooth in ten dimensions, and    
the internal geometry $W_z$ smoothly shrinks at $z=z_\text{max}$, and 
we encounter a ``boundary'' of the geometry only after the description 
on 10 dimensions is reduced to that on the five dimensions.  

In the following, we construct a toy model using the Klebanov-Strassler
metric so that the model captures the above nature of faithful solutions 
to the equation of motions. With this toy model, where the IR geometry is 
treated in a more appropriate way than in the hard wall model, 
numerical calculation is carried out once again, in order to 
check the validity of the crude estimation of the overlap integration 
in section \ref{sec:derivation}, and also to see how much individual 
$(I+2)$-point coupling constants are different from their average. 
Because the Klebanov-Strassler metric does not asymptote to a pure 
$AdS_5 \times W$ metric for some 5-dimensional manifold $W$ 
in the UV region ($z \sim 0$), but maintains logarithmic running, 
our toy model using the Klebanov-Strassler metric also serves 
as an ideal test case of how to deal with the running $a(z)$ function 
in the NDA scaling rule.

The Klebanov-Strassler background is dual to an ${\cal N}=1$ $SU(N)\times SU(N+M)$ quiver gauge theory with chiral multiplets in the bifundamental representations \cite{Klebanov:2000hb}.
This gauge theory experiences a cascade of Seiberg duality, $SU(N)\times SU(N+M)\rightarrow SU(N-M)\times SU(N)$.
In the IR, the duality cascade ends when the gauge group becomes $SU(M)$ and confinement occurs.
In the UV, the cascade of Seiberg duality continues unlimitedly,
and it has weakly running RG flow in the UV; this theory is not UV conformal. 

Let us briefly review the Klebanov-Strassler metric, focusing only on
the aspects that we need in the following. 
The background metric is given by \cite{Klebanov:2000hb},
\begin{align}
 ds_{10}^2 &= h^{-1/2}(\tau)dx^2 + h^{1/2}(\tau)ds_6^2,\label{KS} \\
 ds_6^2 &=\frac{\epsilon^{4/3}}{2}K(\tau)\left[
    \frac{1}{3K^3(\tau)}[(d\tau)^2+(g^5)^2]
   +\cosh^2\sfrac{\tau}{2}[(g^3)^2+(g^4)^2]   \right. \notag \\
  & \qquad \qquad \qquad \qquad \qquad \qquad \qquad \quad \quad 
   \left. +\sinh^2\sfrac{\tau}{2}[(g^1)^2+(g^2)^2]
\right],  
\label{eq:dc}
\end{align}
where $\tau\;(\ge 0)$ is the radial coordinate, $g^{1},\dots, g^{5}$ are basis of 1-forms on the compact five dimensional space, 
and $\epsilon^{2/3}$ is a parameter with mass dimension $-1$.
The two functions appearing in (\ref{KS}, \ref{eq:dc}) are given by
 \begin{align}
 K(\tau)=\frac{(\sinh(2\tau)-2\tau)^{1/3}}{2^{1/3}\sinh(\tau)},\label{eq:K(tau)}
 \quad h(\tau)=\alpha \frac{2^{2/3}}{4}\int_\tau^\infty dx\frac{x\coth
  x-1}{\sinh^2 x} (\sinh (2x)-2x)^{1/3}, 
\end{align}
where $\alpha = 4(g_s M \alpha')^2 \epsilon^{-8/3}$ is a dimensionless constant determined by the dilaton vev, RR 3-form flux $M$ and the parameter $\epsilon^{2/3}$.
We can rewrite the metric (\ref{KS}) into the form 
(\ref{eq:general geometry}), by defining the holographic coordinate $z$
and the function $f(z)$ in (\ref{eq:general geometry}) as 
\begin{align}
 \frac{dz}{d\tau} = - \frac{\epsilon^{2/3}}{\sqrt{6}}\frac{(h(\tau))^{1/2}}{K(\tau)},\quad z(\tau=\infty)=0,\quad f^2(z)=(h(\tau))^{1/2}.
\label{eq:z(tau)}
\end{align}

In the UV limit ($\tau \rightarrow \infty$, $z \rightarrow 0$), 
the asymptotic form of $K(\tau)$ and $h(\tau)$ are given by 
$K(\tau)\rightarrow 2^{1/3}e^{-\tau/3}$ and 
$h(\tau)\rightarrow \alpha 2^{-5/3} 3 \tau e^{-4\tau/3}$, respectively, 
and hence the new holographic coordinate $z$ is related to the original 
one $\tau$ by $z\rightarrow \alpha^{1/2}\epsilon^{2/3}
2^{-5/3}3 \tau^{1/2} e^{-\tau/3}$
 asymptotically. The asymptotic form of
the metric \cite{Klebanov:2000nc} is given (in this coordinate $z$) 
by,\footnote{We follow the definition of $R^2$ in \cite{Herzog:2002ih}.}
\begin{align}
 ds_{10}^2 \rightarrow  \frac{R^2}{z^2}(\ln (z_0/z))^{1/2} (\eta_{\mu\nu} dx^\mu dx^\nu  + dz^2) + R^2 (\ln (z_0/z))^{1/2} ds_{T^{1,1}}^2,\quad R^2= \frac{9 g_s M \alpha'}{2\sqrt{2}},
\label{eq:asymptotic form of KS}
\end{align}
where $ds^2_{T^{1,1}}=\frac{1}{9}(g^5)^2+\frac{1}{6}\sum_{i=1}^4(g^i)^2$
is the metric of a five dimensional compact manifold, known as
$T^{1,1}$, which is topologically $S^2 \times S^3$.
Thus, the Klebanov-Strassler metric is approximately that of 
$AdS_5\times T^{1,1}$ in the UV limit, except for the slowly varying 
logarithmic factors $\ln(z_0/z)$; $z_0$ is a constant of order $\alpha^{1/2}\epsilon^{2/3}$.

In the IR limit ($\tau \rightarrow 0$),
the functions in (\ref{eq:K(tau)}) have finite limits, $K(0)= (2/3)^{1/3}$ and $h(0)\simeq\alpha\times 0.285$. 
The two directions $g^1$ and $g^2$ on the compact space shrink as
$\propto \frac{1}{2}(d\tau)^2 +  \frac{\tau^2}{4}[(g^1)^2 +(g^2)^2]$.
This form of the metric in the $\tau \simeq 0$ region is like a three 
dimensional flat metric written in polar coordinates, implying that 
the geometry ends smoothly at $\tau=0$. The three other directions
spanned by $g_3,~g_4$ and $g_5$ form a three dimensional sphere at
$\tau=0$. Thus, the internal six-dimensional geometry is locally 
$R^3 \times S^3$, and $\tau$ is a radial coordinate of $R^3$ 
\cite{Candelas:1989js}.
In the $z$-coordinate (\ref{eq:z(tau)}), the smooth endpoint of the 
geometry $\tau=0$ corresponds to the maximal value of $z$, 
$z_\text{max}=z(\tau=0)\simeq (\alpha^{1/2}\epsilon^{2/3})\times 1.50$.
The warp factor $(f(z))^{-2}$ has its nonzero minimal value 
$\text{min} [(f(z))^{-2}] = (f(z_\text{max}))^{-2} \simeq (\alpha\times
0.285)^{-1/2}$ at $z=z_{\rm max}$, which can also be written as 
$\text{min}[f(z)]^{-2} \simeq 2.65 \times (R/z_{\rm max})^2$.

Let us now carry out dimensional reduction of Type IIB supergravity
action on the Klebanov-Strassler metric (\ref{eq:dc}), first 
to five dimensions. In non-conformal theories like this, 
the factor $(R^5 {\rm Vol}(W_z)) \times f^{-3}(z)$ most relevant to 
the dimensional reduction of dilaton is not simply proportional to 
$1/z^3$, and it does not even asymptotes $1/z^3$ at small $z$ in 
the UV non-conformal theories. In the Klebanov-Strassler metric,
however, 
\begin{eqnarray}
 \left[(R^5 {\rm Vol}(W_z)) \times f^{-3}(z) \right] 
 & = & \left[ \left( \left(\frac{3}{2}\right)^{5/2} 
                 K(\tau) h^{5/4}(\tau) \sinh^2(\tau) 
                 \times \epsilon^{10/3} {\rm Vol}(T^{1,1}) 
          \right) f^{-3}(z) \right]  \nonumber \\
 & \rightarrow & R^8 \times {\rm Vol}(T^{1,1}) \times 
   \frac{\left( \ln(z_0/z) \right)^2}{z^3}, 
\end{eqnarray}
in the UV region, and it is a reasonable approach to take 
only the $R^8 {\rm Vol}(T^{1,1})$ factor out of the integral, 
just like in (\ref{eq:cftaction}), while keeping $(\ln(z_0/z))^2$ and 
$1/z^3$ inside the integral. The action in five dimensions starts 
with the following terms:
\begin{align}
S=\sfrac{4a_\text{eff}}{8\pi^2} \int d^4x \int^{z_{\rm max}}_0 
  \frac{dz}{z^3}\bar Y(z)
  \left( -\frac{1}{2}((\partial \phi')^2+(\partial_z \phi')^2)-\frac{1}{2} e^{2\phi'}((\partial c')^2+(\partial_z c')^2) \right)+\cdots.
\label{eq:KSaction}
\end{align}
$\phi'(x,z)$ and $c'(x,z)$ are dilaton and RR 0-form
fields,\footnote{The field redefinition just below
(\ref{CFT5dimaction}) does not involve any technical subtleties, 
because the dilaton vev is constant in the Klebanov-Strassler background. }
respectively, with a constant profile over the internal manifold 
$W = T^{1,1}$. 
Here,  
\begin{eqnarray}
&& \bar{Y}(z) = \frac{f^{-3}(z)}{(R/z)^3} 
   \frac{(R^5 {\rm Vol}(W_z))}{R^5 {\rm Vol}(T^{1,1})}, \\
&&   4a_\text{eff}=p' N_\text{eff}^2,\quad N_\text{eff}=\sfrac{3g_s M^2}{2\pi}, \quad p' = \frac{\text{Vol}(S^5)}{\text{Vol}(T^{1,1})}=\frac{27}{16}.
\label{eq:aeff}
\end{eqnarray}
One can see in a straightforward computation that the $a(z)$ function 
defined in (\ref{eq:a(z)}) has an asymptotic form $a(z) \rightarrow
a_{\rm eff} ( \ln (z_0/z))^2$ in the UV (small $z$) region, precisely 
with the coefficient $a_{\rm eff}$ defined above. 

In the rest of this article, we ignore all the terms denoted by 
$+ \cdots$, and take (\ref{eq:asymptotic form of KS}) without the 
``$+\cdots$ terms'' as the starting point of a toy model. 
Non-trivial 3-form flux background $H_3$ and $F_3$
in the Klebanov-Strassler solution would potentially generate potential 
and mixing of $\phi'$ and $c'$ even at this level of description at five 
dimensions, and consequently would affect the mass spectra and coupling 
constants of hadrons in this theory. All these effects are ignored,
however. 
Thus, numerical results from this toy model should not be taken literally 
as results of Klebanov--Strassler solution of string theory. We
primarily use this toy model\footnote{One might also expect that 
the numerically calculated coupling of hadrons of this toy model will 
provide a decent guess of those in the Klebanov-Strassler model within 
an error of a factor of order unity, which is good enough for the
main subject of this article, but we prefer to make an error on a safe
side, and would not push the argument that far in this article.}  
for the purpose we already stated at the beginning of this 
section \ref{ssec:KS}. 
 
\begin{figure}[tbp]
 \begin{center}
\begin{tabular}{ccc}
  \includegraphics[scale=0.8]{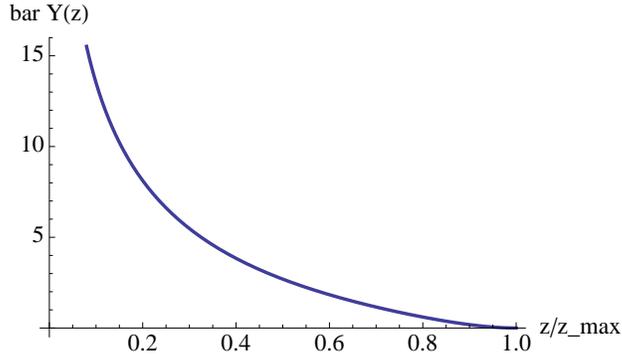} 
    \end{tabular}
 \caption{ $\bar Y(z)$ in (\ref{eq:KSaction}) using the Klebanov-Strassler
metric.
\label{fig:Yzplot} }
 \end{center}
\end{figure}

Now the effects of the confining geometry in the IR region ($z \sim
z_{\rm max}$) and the logarithmic violation of conformal symmetry even 
in UV region $z \rightarrow 0$ are all encoded in the measure function 
$\bar{Y}(z)$ in this model. The profile of $\bar{Y}(z)$ in 
Figure \ref{fig:Yzplot} is neither constant (as in conformal theories)
nor approaches a constant value in the UV ($z\rightarrow 0$) limit,
reflecting the nature of the Klebanov-Strassler metric (\ref{eq:dc}). 
Although the measure $\bar{Y}(z)$ of the overlap integration diverges 
as $(\ln (z_0/z))^2$ in the UV region $z\rightarrow 0$, and vanishes 
as $\propto (z-z_{\rm max})^2$ at the IR boundary, its value remains 
quite stable and moderate, $4\sim 1$, for a middle range 
$0.4 \lesssim z/z_\text{max} \lesssim 0.8$ in the holographic radius.
This is where dominant contribution to the overlap integrals comes from, 
and hence there is no reason to suspect that the discussion 
in section \ref{sec:derivation} might go wrong in this case; 
in evaluating $a(z)$ at $z \sim \gamma z_{\rm max}$ to be used 
as $\bar{a}$ in the NDA scaling factor (\ref{Eq:our statement2}), 
we only need to take $\gamma$ in this middle range, $0.4\sim 0.8$.

Numerical calculation of hadron couplings in {\it this} model also shows 
that the $(I+2)$-point coupling constants follow the scaling rule, when 
the scaling factor is chosen to be
\begin{equation}
 \left(\frac{4\pi}{\sqrt{2a_{\rm eff}}} \Lambda_{\rm NDA}z_{\rm max}
  \cdot 1.5 \right).
\end{equation}
See Table \ref{KScoupl}. 
Because $\bar{a}$ evaluated somewhere in the middle range 
$\gamma \in [0.4, 0.8]$ of the holographic radius is much the same as
$a_{\rm eff}$ numerically, the crude (and model independent) estimation 
of the overlap integral in section \ref{sec:derivation} holds true quite 
accurately for this model. 

It is also worthwhile to note, just like in section \ref{ssec:HW}, 
that the factor $(\Lambda_{\rm NDA} z_{\rm max})$ becomes $6.2$, if 
the lightest hadron mass eigenvalue $m_1$ is used for 
$\Lambda_{\rm NDA}$ in this toy model. The NDA scaling rule holds, 
but the scaling factor, $(4\pi) \times 9$ apart from the degree of
freedom factor $\sqrt{2a}$, is larger than expected in the NDA ansatz.

\begin{table}[h]
 \caption{\label{KScoupl} 
Geometric means
$a^{(I)}_\text{typ}$ and  $b^{(I)}_\text{typ}$
and the standard deviations of $\ln |a^{(I)}_{n_1 n_2 m_1 \dots m_I}|$ and
$\ln |b^{(I)}_{n_1 n_2 m_1\dots m_I}|$, $\sigma^{(I)}_{\ln a}$ and 
$\sigma^{(I)}_{\ln b}$, respectively, calculated numerically in 
a model (\ref{eq:KSaction}) using the Klebanov-Strassler metric.
Just like in Table \ref{HWresult}, $\sigma^{(I)}$ in the 3rd and 5th
columns are presented in the form of $\ln[ \; \exp(\sigma^{(I)}) \; ]$.
}
\begin{center}
\begin{tabular}[c]{|c|c|c|c|c|}
 \hline 
 &
\hspace{-1.8mm}$\displaystyle{a^{(I)}_\text{typ}\left(\frac{4\pi}{\sqrt{2a_\text{eff}}} \Lambda_\text{NDA}z_\text{max} \cdot 1.5 \right)^{-I}}$\hspace{-1mm}
 &
\hspace{-1mm} $\sigma^{(I)}_{\ln a}$ 
 &
 $\hspace{-1.8mm}\displaystyle{b^{(I)}_\text{typ}\left(\frac{4\pi}{\sqrt{2a_\text{eff}}}
	     \Lambda_\text{NDA}z_\text{max} \cdot 1.5
	     \right)^{-I}}$\hspace{-1mm}&
  $\sigma^{(I)}_{\ln b}$ 
\\
\hline
$I=1$ &
0.3 &
$\ln [2.]$
&
$\sfrac{m_1}{\Lambda_\text{NDA}}^2 \times 0.3$
&
$\ln [3.]$
\\
$I=2$ &
0.3 &
$\ln [1.]$
&
$\sfrac{m_1}{\Lambda_\text{NDA}}^2 \times 0.2$
&
$\ln [4.]$
\\
$I=3$ &
0.4 &
$\ln [2.]$
&
$\sfrac{m_1}{\Lambda_\text{NDA}}^2 \times 0.2$
&
$\ln [4.]$
\\
$I=4$ &
0.5 &
$\ln [2.]$
&
$\sfrac{m_1}{\Lambda_\text{NDA}}^2 \times 0.2$
&
$\ln [3.]$
\\\hline

\end{tabular}
\end{center}
\end{table}

\section{Summary}

Our study based on gauge/string duality is summarized as follows:
the NDA scaling rule does exist for the scalar glueballs dual to $\phi$ and $c$ fields, and the scaling factor is given by (\ref{Eq:our statement2}).
The error of the scaling factor is within a factor two or so (0.8 for hard wall model and 1.5 for the model using the Klebanov-Strassler metric.) 
The uncertainty in the choice of $a_\text{eff}$ is also of the same
order. The $N_c$ dependence of hadron couplings following from the 
large $N_c$ argument (and from string theory) is now generalized 
in the language of $a(z)$ function defined in holographic models, 
which roughly characterizes the degree of freedom in the dual theories. 
With this generalization, the NDA scaling rule can be applied also 
to some class of quiver gauge theories. 

If we are to use the mass of the lightest (non-Nambu--Goldstone boson) 
hadron as the $\Lambda_{\rm NDA}$ mass scale in (\ref{NDA_ansatz}), 
the scaling factor comes out to be larger than the conventional one, 
by $(m z_{\rm max})$, which is about $5\sim 6$ in the two models we
studied numerically. We can say that the (dimensionless) coefficients of
hadron couplings are systematically larger than the conventional NDA
ansatz, but alternatively, we can also say, by taking $\Lambda_{\rm NDA}
\simeq 1/z_{\rm max}$, that the coefficients of hadron couplings are 
just as expected in the conventional NDA ansatz, and mass parameters 
are somewhat larger than naively expected from $\Lambda_{\rm NDA}$.
We have nothing to say, however, about whether the large value of 
$mz_{\rm max}$ is a generic feature of hadrons from strongly coupled 
theories, or just happens to be a specific feature of the two models 
we studied. 

So far, we are not sure whether coupling constants of any other hadrons have
the NDA scaling rule, and if so, what are the NDA scaling factors.
These are subjects in our future work.

\section*{Acknowledgments}
We thank Y.~Tachikawa for sharing his knowledge on the volumes of 
Einstein manifolds, and S.~Sugimoto and W.~Yin for useful discussions.
The work is supported in part by JSPS Research Fellowships for 
Young Scientists (RN and KY), 
World Premier International Research Center Initiative
(WPI Initiative), MEXT, Japan (RN, TW, TTY, KY), 
a Grant-in-Aid for Scientific Research on Innovative 
Areas 2303 (TW), and by 
Grant-in-Aid for Scientific research from the MEXT, Japan, No. 22244021 (TTY).

\providecommand{\href}[2]{#2}
\begingroup
\raggedright

\end{document}